\documentclass[useAMS]{mn2e}
\usepackage{graphicx}
\usepackage{epsfig}
\usepackage{epsf}
\usepackage{amsmath}
\usepackage{amssymb}
\usepackage{stfloats}
\title[SN2011ht]{SN~2011ht: Confirming a Class of Interacting Supernovae with Plateau Light Curves (Type IIn-P)}
\author[Mauerhan et al. ]{Jon C. Mauerhan$^{1}$\thanks{E-mail:
mauerhan@as.arizona.edu}, Nathan
Smith$^{1}$, Jeffrey M. Silverman$^{2}$, Alexei V. Filippenko$^{2}$,  \newauthor  Adam N. Morgan$^{2}$, S. Bradley Cenko$^{2}$, Mohan Ganeshalingam$^{2}$, Kelsey I. Clubb$^{2}$,  \newauthor Joshua S. Bloom$^{2}$, Thomas Matheson$^{3}$, and Peter Milne$^{1}$  \\
$^{1}$University of Arizona, Steward Observatory, Tucson, Arizona  85721, USA \\ $^2$Department of Astronomy, University of California, Berkeley, CA 94720-3411, USA \\ $^{3}$National Optical Astronomy Observatory, Tucson, AZ 85719, USA}
\begin{document}


\pagerange{\pageref{firstpage}--\pageref{lastpage}} \pubyear{2012}

\maketitle

\label{firstpage}

\begin{abstract}
We present photometry and spectroscopy of the Type IIn supernova (SN) 2011ht, identified previously as a possible SN impostor. The light curve exhibits an abrupt transition from a well-defined $\sim 120$ day plateau to a steep bolometric decline, plummeting 4--5 mag in the optical and 2--3 mag in the infrared in only $\sim 10$ days.  Leading up to peak brightness ($M_V=-17.4$ mag), a hot emission-line spectrum exhibits strong signs of interaction with circumstellar material (CSM), in the form of relatively narrow P-Cygni features of H~{\sc i} and He~{\sc i} superimposed on broad Lorentzian wings. For the latter half of the plateau phase, the spectrum exhibits strengthening P-Cygni profiles of Fe~{\sc ii}, Ca~{\sc ii}, and H$\alpha$.  By day 147, after the plateau has ended, the SN entered the nebular phase, heralded by the appearance of forbidden transitions of [O~{\sc i}], [O~{\sc ii}], and [Ca~{\sc ii}] over a weak continuum. At this stage, the light curve exhibits a low optical luminosity that is comparable to that of the most subluminous Type II-P supernovae, and a relatively fast visual wavelength decline that appeared to be significantly steeper than the $^{56}$Co decay rate. However, the total pseudo-bolometric decline, including the infrared luminosity, is consistent with $^{56}$Co decay, and implies a low $^{56}$Ni mass in the range  $0.006$--$0.01\, {\rm M}_{\odot}$, near the lower end of the range exhibited by SNe~II-P.  We therefore characterize SN~2011ht as a core-collapse SN very similar to the peculiar SNe~IIn 1994W and 2009kn. These three SNe appear to define a subclass, which are Type IIn based on their spectrum, but that also exhibit well-defined plateaus and produce low $^{56}$Ni yields. We therefore suggest Type IIn-P as a name for this subclass. The absence of observational signatures of high-velocity material from SNe IIn-P could be the result of an opaque shell at the shocked SN-CSM interface, which remains optically thick longer than the time scale for the inner ejecta to cool and become transparent. Possible progenitors of SNe~IIn-P, consistent with the available data, include 8--10 \,M$_{\odot}$ stars, which undergo core collapse as a result of electron capture after a brief phase of enhanced mass loss, or more massive ($M\gtrsim25\, {\rm M}_{\odot}$) progenitors, which experience substantial fallback of the metal-rich radioactive ejecta. In either case, the energy radiated by these three SNe during their plateau (2--$3\times10^{49}$ erg for SN~2011ht) must be dominated by CSM interaction, and the subluminous tail the result of low $^{56}$Ni yield.
\end{abstract}


\section{Introduction}
Type~IIn supernovae (SNe~IIn) are a subset of core-collapse events that exhibit relatively narrow emission lines in their spectra (Schlegel 1990; Filippenko 1997), indicating the presence of dense circumstellar material (CSM) that envelops the SN explosion (Chugai 1990).  This CSM, which masks the broad emission and absorption features typically seen in normal SNe, must have been ejected by the progenitor in the years to decades prior to explosion. The enhanced mode of pre-SN mass loss can occur over a large range of time scales, from eruptive events months to years before the explosion, to long-duration superwinds that blow for millennia.  As such, multi-epoch observations of SNe~IIn allow us to probe the mass-loss parameters and physical states of massive stars immediately preceding their core collapse, providing a unique glimpse into the final phases of their evolution. 

The progenitors of SNe~IIn may span the entire range of massive-star classes, including luminous blue variables (LBVs; e.g., SN 2005gl, Gal-Yam \& Leonard 2009; SN 2006tf, SN 2006gy, Smith et al. 2007, 2008; SN~1961V, Smith et al. 2011a, Kochanek et al. 2011; SN~2009ip, Mauerhan et al. 2012), red supergiants (RSGs; e.g., SN 2005ip, Smith et al. 2009; SN 1998S, Bowen et al. 2000, Mauerhan \& Smith 2012), Wolf-Rayet stars (e.g., SN 1985F; Filippenko \& Sargent 1985, Begelman \& Sarazin 1986), as well as transitional Ofpe/WN9 stars on their way to the Wolf-Rayet phase (e.g., Smith et al. 2012a). Even the least massive of stars that are capable of becoming core-collapse SNe (8--10\,M$_{\odot}$) might potentially generate a Type~IIn explosion, a possibility considered in the case of the SN~IIn 1994W (Sollerman et al. 1998). Thus, a diverse range of progenitors can potentially experience a highly enhanced degree of mass loss prior to exploding, but the time scales involved may vary significantly for evolved massive stars. Indeed, the physical mechanisms responsible for triggering the pre-SN superwinds and LBV eruptions are not understood. The fact that SNe~IIn comprise only $\sim 9$\% of all core-collapse events (Smith et al. 2011b; Li et al. 2011) complicates this picture, as there is no clear explanation why only a small fraction of massive stars spanning a large progenitor mass range experience such a highly enhanced degree of mass loss immediately before exploding.

For SNe~IIn, the magnitude of the emission from CSM interaction is sensitive mainly to the velocity of the SN shock and the pre-SN wind density. These parameters can span a broad range, depending on the nature of the progenitor and its mass-loss history. It is thus difficult to gauge the magnitude of radioactive-element synthesis, or lack thereof, until after CSM interaction has diminished substantially. Some SNe~IIn, such as the exceptionally luminous SN 2006gy, maintain powerful CSM interaction for so long that any radioactive-decay emission would fade away before becoming discernible in the light curve. Other interacting SNe, such as SNe 1980K and 1998S, reveal the common radioactive light-curve slope at earlier times, but still maintain enough CSM emission at late times to complicate or prevent radioactive diagnosis (e.g., Milisavljevic et al. 2012; Mauerhan \& Smith 2012). Finally, a rare subset of SNe~IIn that exhibit plateaus in their light curves, similar in shape to those of SNe~II-P, become exceptionally faint after the plateau, well below the typical luminosity of radioactive decay. The SN~IIn 1994W is of this latter subset, exhibiting an exceptionally faint and steeply declining light curve during its nebular phase, after a well-defined plateau. 

Multiple interpretations have been considered in attempting to discern the peculiar properties of SN~1994W, including explosive pre-SN mass loss (Chugai et al. 2004) that was followed by an explosion having a low $^{56}$Ni yield (Sollerman et al. 1998), as well as a non-SN interpretation of colliding stellar mass-loss shells (Dessart et al. 2009). Few SNe have exhibited the unusual combination of observational properties displayed by SN~1994W. However, the SN~IIn 2009kn was recently considered by Kankare et al. (2012) to be a ``twin" of SN~1994W, based on the spectral morphology and a very similar well-defined plateau light curve. More recently, the discovery of SN~2011ht (Roming et al. 2011) has shown this SN to be an even closer twin of SN~1994W (Roming et al. 2012; Humphreys et al. 2012), exhibiting strikingly similar spectral evolution, a well-defined $\sim 110$--120 day plateau, and an exceptionally faint and steeply declining optical decay tail. In addition, the IIn SN~2005cl (Kiewe et al. 2012) shares the same spectral characteristics and well-defined plateau as SN ~2011ht, SN~1994W, and SN~2009kn, although the plateau appears significantly more luminous. The late time luminosity of SN~2005cl was not constrained, however. 

Soon after discovery, SN~2011ht was identified as an unusual object. Initially, it was classified by Roming et al. (2011, ATEL 3690) as a possible SN  ``impostor" --- the eruption of an LBV. This explosion was one of the first to be observed in the UV band during the rise to peak, revealing an extreme UV brightening of $\sim 7$ mag, compared to $\sim 2$ mag in the optical, which probably resulted from shock-breakout photons degraded by diffusion through dense CSM. X-ray emission detected with \textit{Swift} was also reported by Roming et al. (2012), although subsequent high-resolution X-ray imaging with \textit{Chandra} revealed that this was likely to be a false match with a background X-ray source (Pooley 2012, ATEL 4062). The peak absolute visual magnitude of $\sim -17$, and early spectral evolution led to the conclusion that SN~2011ht was perhaps a true SN IIn, not an impostor (Prieto et al. 2011, CBET 2903). In further support of this interpretation, the total radiated energy of $2.5\times10^{49}$ erg derived for SN~2011ht during the first 110 days is far more luminous than typical LBV eruptions,  unless the latter have a much wider luminosity distribution than currently thought (see Smith et al. 2011a). Still, Humphreys et al. (2012) have continued to favor the SN impostor hypothesis for SN~2011ht. 

Here, we present photometric and spectroscopic observations of SN~2011ht. Based on the following analysis and comparison with SN~1994W, SN~2009kn, and, to a lesser extent, SN~2005cl, we interpret SN~2011ht as a true SN, and not a SN impostor. As such, it helps solidify a new class of interacting SNe that exhibit well-defined plateaus, spectra dominated by CSM interaction, and faint decay tails. 

\section{Observations}

\begin{table}\begin{center}\begin{minipage}[bp]{3.2in} \setlength{\tabcolsep}{5.5pt}
\caption{Optical photometry of SN~2011ht. Epochs are given with respect to discovery date (2011 Sep. 29). The uncertainties represent the standard deviation of the zero-point magnitudes derived from measurements of 4--5 field stars.} \scriptsize
\centering
  \begin{tabular}[bp]{@{}lcccc@{}} 
  \hline
   JD$-$2,450,000      & $B$ & $V$ & $R$ & $I$   \\                            
   /Epoch (days)       & (mag) & (mag) & (mag) & (mag) \\
\hline
\hline
5868.8/35 & $14.41(0.05)$ &  $14.46(0.14)$ & $14.23(0.09)$
& $14.18(0.08)$ \\
5873.8/40 & $14.32(0.05)$ &  $14.37(0.14)$ & $14.15(0.09)$
& $14.08(0.07)$ \\
5875.8/42 & $14.26(0.06)$ &  $14.38(0.15)$ & $14.16(0.12)$
& $14.13(0.11)$ \\
5879.5/46 & $14.28(0.05)$ &  $14.35(0.15)$ & $14.15(0.13)$
& $14.14(0.15)$ \\
5888.8/55 & $14.50(0.10)$ &  $14.26(0.17)$ & $14.17(0.14)$
& $14.16(0.14)$ \\
5893.5/60 & $14.37(0.06)$ &  $14.45(0.21)$ & $14.24(0.18)$
& $14.18(0.17)$ \\
5898.5/65 & $14.51(0.06)$ &  $14.59(0.21)$ & $14.39(0.19)$
& $14.21(0.21)$ \\
5901.5/68 & $14.52(0.06)$ &  $14.57(0.21)$ & $14.34(0.17)$
& $14.23(0.17)$ \\
5905.5/72 & $14.60(0.06)$ &  $14.64(0.21)$ & $14.32(0.15)$
& $14.24(0.17)$ \\
5912.5/79 & $14.71(0.05)$ &  $14.69(0.15)$ & $14.46(0.14)$
& $14.35(0.13)$ \\
5924.8/91 & $15.07(0.06)$ &  $14.95(0.21)$ & $14.68(0.19)$
& $14.52(0.20)$ \\
5933.5/100 & $15.46(0.04)$ &  $15.20(0.20)$ & $14.85(0.20)$
& $14.78(0.21)$ \\
5936.5/103 & $15.51(0.06)$ &  $15.23(0.21)$ & $14.91(0.19)$
& $14.69(0.19)$ \\
5943.5/110 & $15.80(0.06)$ &  $15.35(0.21)$ & $15.03(0.19)$
& $14.79(0.20)$ \\
5957.8/124 & $16.68(0.05)$ &  $15.88(0.15)$ & $15.37(0.11)$
& $15.03(0.12)$ \\
5961.5/128 & $17.38(0.06)$ &  $16.49(0.21)$ & $15.94(0.19)$
& $15.55(0.18)$ \\
5974.5/141 & $20.39(0.21)$ &  $19.96(0.20)$ & $19.11(0.20)$
& $18.48(0.21)$ \\
6036.5/203 & ... & 21.39(0.25) & 20.77(0.20) & 20.37(0.15) \\
6054.6/221 & ... & ... & 21.43(0.15) & ... \\

\hline
\end{tabular} \end{minipage} \end{center}
\end{table}

\begin{table}\begin{center}\begin{minipage}{3.2in} \setlength{\tabcolsep}{5.5pt}
\caption{PAIRITEL near-IR photometry of SN~2011ht. The uncertainties are statistical.} \scriptsize
  \begin{tabular}{@{}lccc@{}}
  \hline
   JD$-$2,450,000 & $J$ & $H$ & $K$   \\                            
   /Epoch (days) & (mag) & (mag) & (mag) \\
               \hline
               \hline
5861.5/28 & $14.23(0.03)$ & $13.97(0.05)$ & $13.72(0.08)$ \\
5864.5/31 & $14.05(0.02)$ & $13.86(0.03)$ & $13.61(0.06)$ \\
5866.5/33 & $14.00(0.02)$ & $13.88(0.03)$ & $13.67(0.09)$ \\
5882.5/49 & $13.92(0.02)$ & $13.78(0.03)$ & $13.51(0.05)$ \\
5888.4/55 & $13.89(0.02)$ & $13.68(0.03)$ & $13.43(0.05)$ \\
5892.4/59 & $13.88(0.03)$ & $13.70(0.05)$ & $13.62(0.09)$ \\
5894.4/61 & $13.86(0.02)$ & $13.70(0.03)$ & $13.48(0.07)$ \\
5900.5/67 & $13.94(0.02)$ & $13.75(0.03)$ & $13.62(0.04)$ \\
5903.5/70 & $13.94(0.02)$ & $13.77(0.02)$ & $13.54(0.04)$ \\
5922.4/89 & $14.09(0.02)$ & $13.94(0.03)$ & $13.65(0.06)$ \\
5928.4/95 & $14.11(0.02)$ & $13.92(0.02)$ & $13.63(0.06)$ \\
5932.4/99 & $14.17(0.02)$ & $13.94(0.03)$ & $13.70(0.06)$ \\
5936.4/103 & $14.21(0.03)$ & $13.94(0.05)$ & $13.82(0.14)$ \\
5949.4/116 & $14.34(0.02)$ & $14.07(0.03)$ & $13.83(0.06)$ \\
5953.4/120 & $14.33(0.02)$ & $14.10(0.05)$ & $13.97(0.09)$ \\
5959.4/126 & $14.66(0.02)$ & $14.37(0.03)$ & $14.17(0.06)$ \\
5961.4/128 & $15.01(0.06)$ & $14.76(0.11)$ & $14.61(0.23)$ \\
5964.3/131 & $16.04(0.07)$ & $15.45(0.09)$ & ::: \\ 
5966.3/133 & $16.79(0.07)$ & $16.59(0.12)$ & ::: \\
5968.4/135 & $16.77(0.10)$ &  ::: & ::: \\
5977.4/144 & $18.28(0.25)$ & $17.03(0.20)$ & $16.28(0.2)$ \\
\hline
\end{tabular} \end{minipage} \end{center}
\end{table}

\begin{table}\begin{center}\begin{minipage}{3.2in} \setlength{\tabcolsep}{5.5pt}
\caption{Spectroscopic observations of SN~2011ht} \scriptsize
  \begin{tabular}{@{}lccc@{}}
  \hline
  JD$-$2,450,000 & Facility & Coverage & $R$   \\                            
  /Epoch (days) & & (\AA) & $\delta\lambda/\lambda$   \\
               \hline
               \hline
5860/26 & Lick/Kast  & 3436--9920 & 1400, 900\\
5867/33 & Keck/LRIS &3362--5630 & 1400 \\
5867/33 & Keck/LRIS & 5740--7390 & 2200 \\
5892/58 & Lick/Kast & 3436--9920 & 1400, 900  \\
5898/64 & Keck/LRIS &3362--5630 & 1400 \\
5898/64 & Keck/LRIS & 5740--7390 & 2200 \\
5914/80 & Lick/Kast  & 3436--9920 & 1400, 900 \\
5921/88 & Keck/LRIS &3362--5630 & 1400 \\
5921/88 & Keck/LRIS & 5740--7390 & 2200 \\
5922/89 & Lick/Kast & 3436--9920  & 1400, 900 \\
5928/96 & MMT/BlueCh & 5550--7500 & 4500 \\
5929/96 & Lick/Kast  & 3436--9920 & 1400, 900 \\
5944/111 & Lick/Kast & 3436--9920 & 1400, 900 \\
5946/113 & MMT/BlueCh & 5550--7500 & 4500 \\
5956/123 & MMT/BlueCh & 5550--7500 & 4500 \\
5958/125 & Lick/Kast & 3436--9920 & 1400, 900 \\
5961/127 & KPNO/RCSpec &  3600--8270 &  1100 \\
5981/147 & Lick/Kast & 3436--9920 & 1400, 900 \\
5981/147 & Keck/LRIS &3362--5630 & 1400 \\
5981/147 & Keck/LRIS & 5740--7390 & 2200 \\
5988/155 & MMT/BlueCh & 5550--7500 & 4500 \\ 
6034/201 & MMT/BlueCh & 3920--8998 & 500 \\ 
6036/202 & Lick/Kast & 3436--9920 & 1400, 900 \\
6046/213 & Keck/LRIS & 5740--7390 & 2200 \\
6071/238& MMT/BlueCh & 5550--7500 & 4500 \\ 
\hline
\end{tabular} \end{minipage} \end{center}
\end{table}

We began photometric monitoring of SN~2011ht roughly 28 days after its discovery date of 2011 Sep. 29. We obtained optical $BVRI$  photometry using the 1\,m Nickel telescope at Lick Observatory, and infrared (IR) $JHK$ photometry using the 1.3\,m Peters Automated Infrared Imaging Telescope (PAIRITEL)\footnote{See http://www.pairitel.org/ .} on Mt. Hopkins (Bloom et al. 2006). The photometric measurements were obtained over 17--19 separate epochs between days 28--150 after discovery. Late-time photometry on days 203 and 221 was also obtained using the 90Prime Imager on the Bok 90$^{\prime\prime}$ telescope on Kitt Peak, and the MONT4K Imager on the Kuiper 61$^{\prime\prime}$ telescope on Mt. Bigelow, respectively. A MONT4K image of the SN is shown in Figure 1.The Lick and PAIRITEL photometry was extracted using standard aperture-photometry techniques and calibrated by photometry of 4--6 field stars in the same image as the SN, while the late-time 90Prime and MONT4K photometry was extracted via point-spread-function (PSF) fitting with the IDL Starfinder code, in which case the host galaxy was modeled as local background, allowing for a more precise determination of the faint SN flux. Tables~1 and 2 list the results.

Our spectroscopic monitoring campaign of SN~2011ht utilized the Kast spectrograph (Miller \& Stone 1993) on the 3 m Shane reflector at Lick Observatory, the Low Resolution Imaging Spectrometer (LRIS; Oke et al. 1995) on the Keck I 10 m telescope, the Bluechannel spectrograph on the Multiple Mirror Telescope (MMT), and the RC Spectrograph (RCSpec\footnote{http://www.noao.edu/kpno/manuals/rcspec/rcsp.html}) on the Mayall 4 m reflector at Kitt Peak National Observatory (KPNO).  The spectroscopic observations are summarized in Table~3. The Lick/Kast spectra have moderate resolution ($R \equiv \delta \lambda / \lambda \approx 900$ and 1400 for the red and blue sides of the Kast spectra, respectively) and covered a large wavelength range of 3436--9920\,{\AA}. The Keck/LRIS and MMT/Bluechannel spectra have higher spectral resolving power of $R\sim 2200$ and 4500, respectively. The RCSpec spectra cover a wavelength range of 3200--8270\,{\AA} with a spectral resolving power of $\sim$1100. All spectra were generally obtained at low airmass or with an atmospheric dispersion corrector; otherwise, observations were performed with the slit at the parallactic angle in  order to minimize chromatic differential slit losses (Filippenko 1982). Our data reduction, wavelength and flux calibration followed standard techniques as described by Silverman et al. (2012). Lick/Kast spectra of SN~2009kn, acquired and reduced utilizing the same instruments setting and reduction techniques described above, were also obtained on 2009 Nov.~10 and Dec.~9. Finally, we also reproduce spectra of SN~1994W, originally presented in Chugai et al. (2004).

\begin{figure}
\includegraphics[width=3.3in]{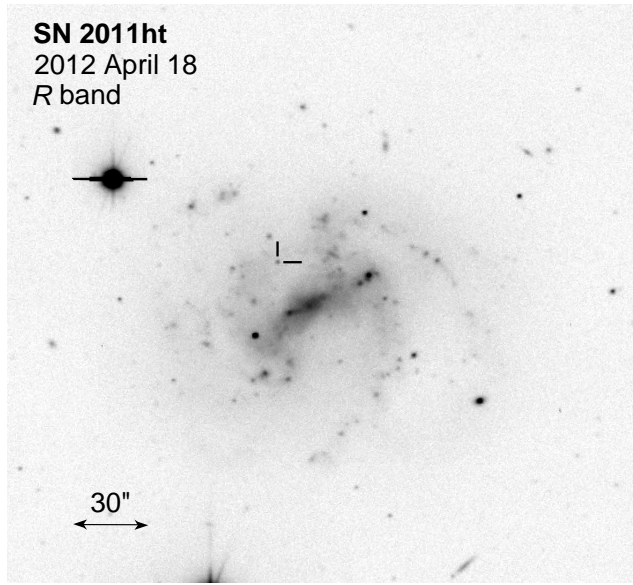}
\caption{$R$-band image of UGC 4560 and SN~2011ht obtained with the 90Prime Imager on 2012 April 18 when the SN was at $R=20.77$ mag. North is up and east to the left.}
\label{fig:finder}
\end{figure}

\begin{figure}
\includegraphics[width=3.3in]{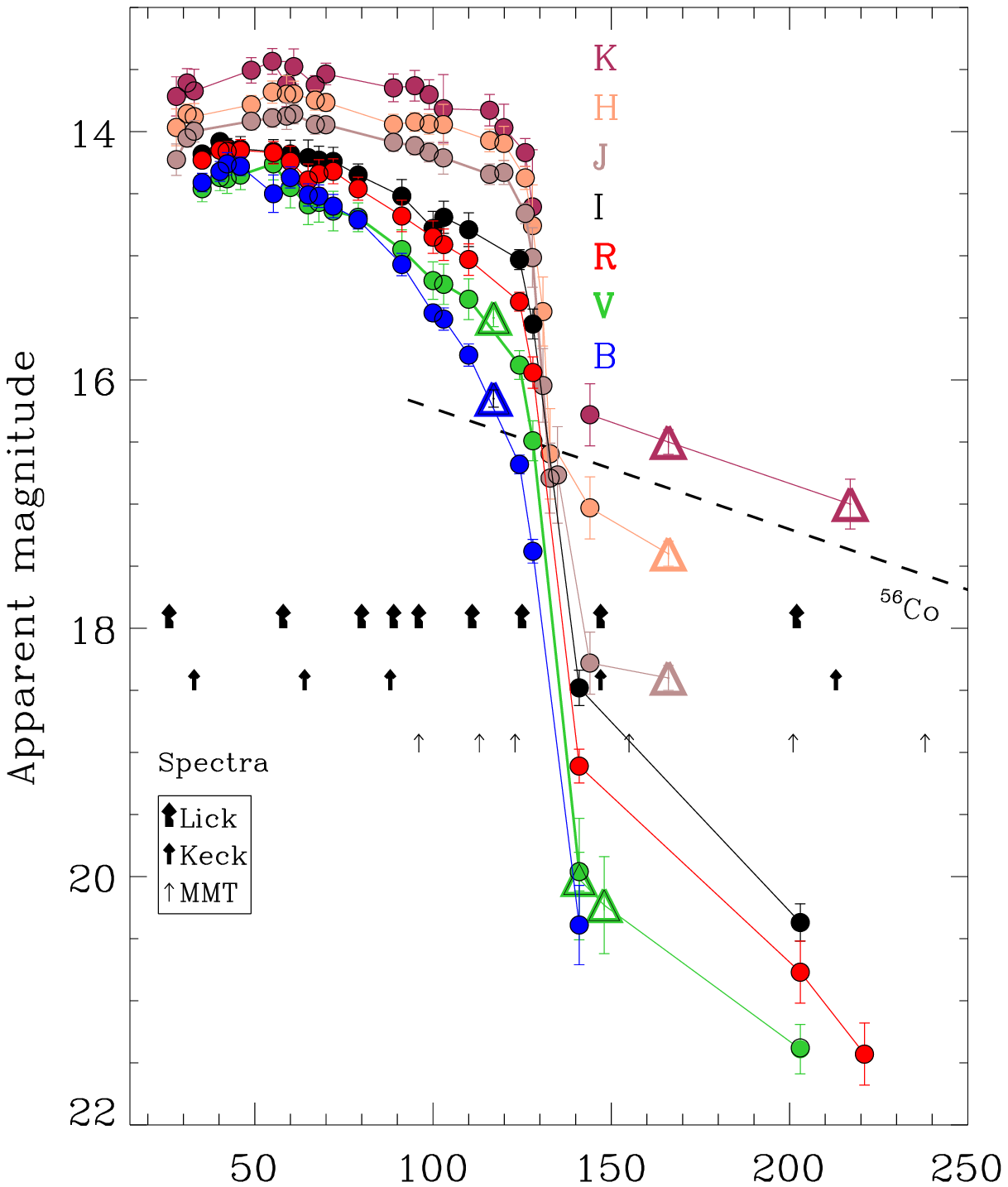}
\includegraphics[width=3.3in]{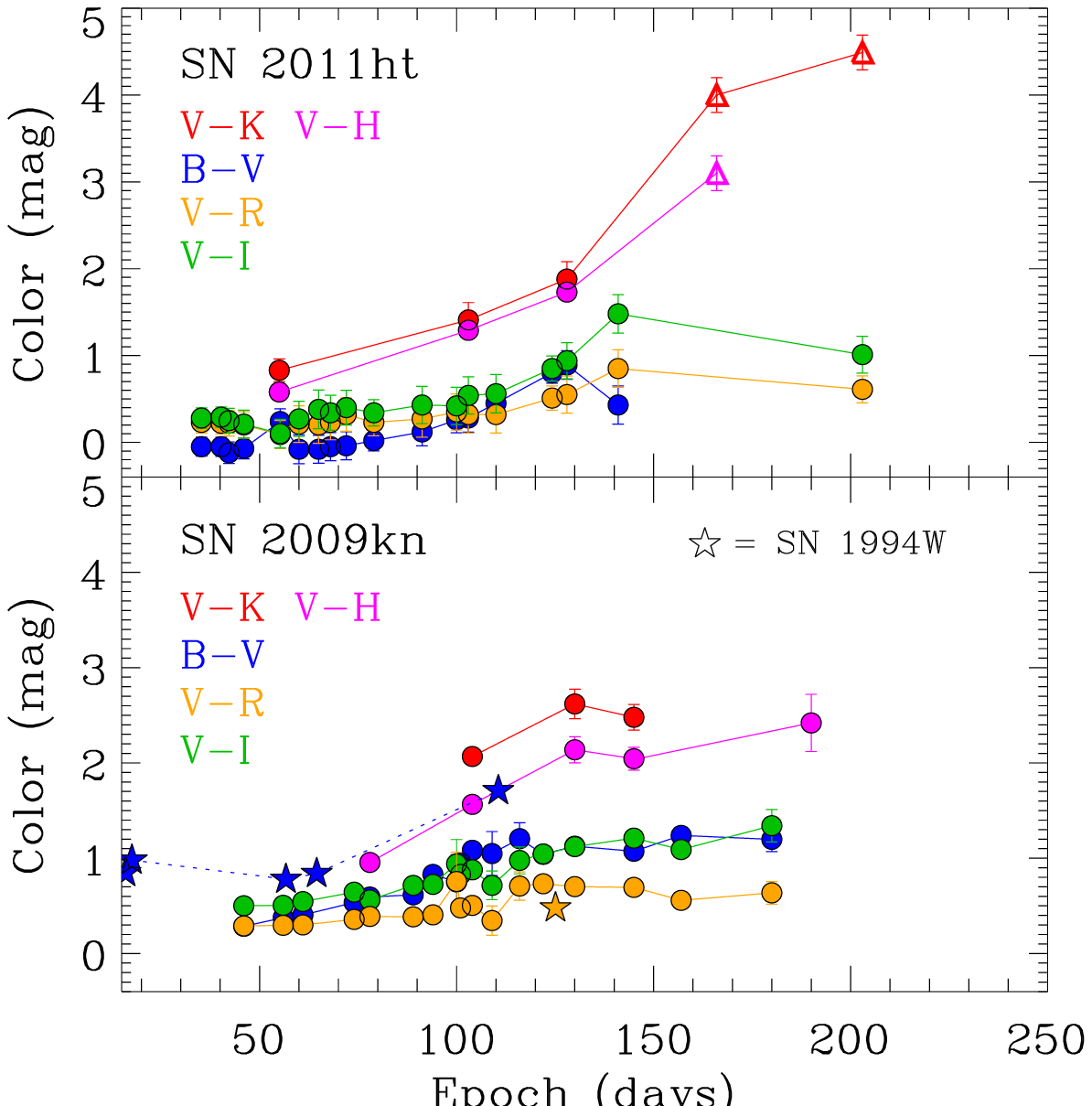}
\caption{$BVRIJHK$ light and color curves of SN~2011ht, with supplemental data from Humphreys et al. (2012; \textit{triangles}). The color curves of SN~2009kn and SN~1994W (from Kankare et al. 2012) have been included for comparison. Epochs having accompanying spectra are marked with upward arrows. The dashed line exhibits the decline rate of $^{56}$Co decay, for an arbitrary $^{56}$Ni mass.}
\label{fig:sn_lc}
\end{figure}

\section{Results}
\subsection{The Light Curve}
The optical and IR light curves of SN~2011ht are presented in Figure~\ref{fig:sn_lc}. The first 20--30 days of monitoring cover the rise to maximum light followed by an extended 120-day plateau phase. The optical light curves peak near day 45 and the IR curves peak approximately two weeks later, near day 60.  Between days 60 and 120, the optical curves begin a shallow decline that steepens with decreasing wavelength. In the $B$ band, the curve gradually drops $\sim 2$ mag before the end of the plateau on day $\sim 120$, while the $JHK$ curves maintain relatively constant brightness for the duration of the plateau. By day 130, the light curve has begun a rapid descent of 4--5 mag in the optical and 2--3 mag in the infrared over a time scale of $\sim 10$ days. By day $\sim 140$, a floor has been reached and the light curves subsequently exhibit a steady decline that continues through at least day 221. 

The color curves at the bottom of Figure~\ref{fig:sn_lc} illustrate the gradual intrinsic reddening of the SN during the plateau phase, as the flux at shorter wavelengths decreases the fastest. Once the plateau phase has ended, the optical colors (including $V-I$) flatten out, becoming slightly bluer. However, there is a substantial shift of flux toward the IR during the nebular phase, as evidenced by the continually increasing $V-H$ and $V-K$ curves. SN~2009kn and SN~1994W exhibit similar color trends in the optical and IR (Kankare et al. 2012), although the late-time IR coverage is not as complete as it is for SN~2011ht. 

\begin{figure}
\includegraphics[width=3.3in]{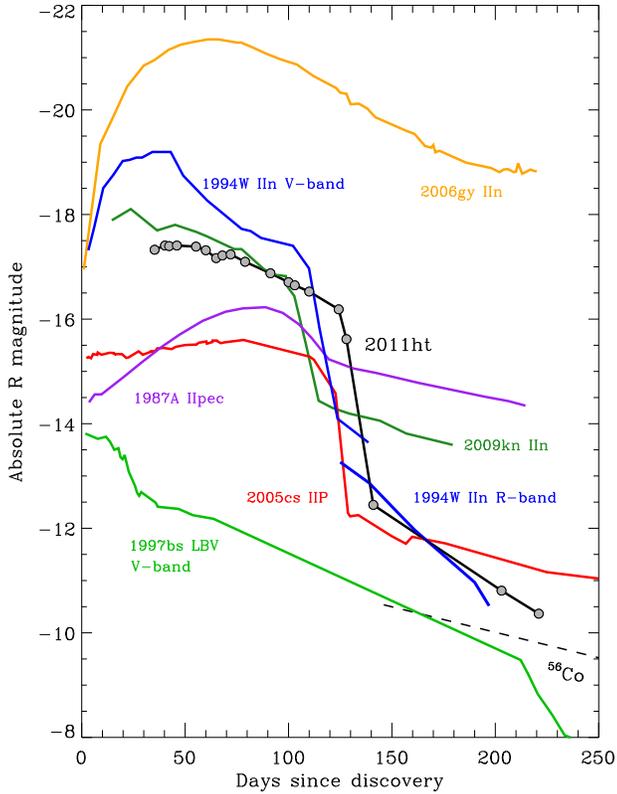}
\caption{Absolute light curve of SN~2011ht, including comparison SNe 1994W (IIn; Sollerman et al.1998), 2009kn (IIn; Kankare et al. 2012), 2005cl (Kiewe et al. 2012); 2005cs (II-P, Pastorello et al. 2009), 1987A (II-pec, Hamuy et al. 1990), and 2006gy (Smith et al. 2007), as well as LBV SN ``impostor" 1997bs (Van Dyk et al. 2000). The height of the SN~2011ht plateau is typical of SNe~IIn, but the late-time tail appears significantly subluminous and declines at a relatively rapid rate, like SN~1994W. The dashed line exhibits the decline rate of $^{56}$Co decay, for an arbitrary $^{56}$Ni mass.}
\label{fig:lc_abs}
\end{figure}

We computed the absolute magnitude of SN~2011ht, adopting an extinction value of $A_V=0.19$ mag and a distance of 19.2 Mpc to the host galaxy UCG 5460 (Roming et al. 2012), which yielded values of $M_V=-17.35$ and $M_R=-17.41$ mag at the respective peaks of days 55 and 46.  The absolute magnitude light curve of SN~2011ht is presented in Figure~\ref{fig:lc_abs}, along with the those of various other core-collapse SNe and a SN impostor for comparison. The well-defined steep-edge plateau of SN~2011ht closely resembles that of the SNe~IIn 2009kn, SN~2005cl, and SN~1994W, although the plateau peaks of SN~1994W and SN~2005cl appeared up to 1.5--2 mag brighter. The plateaus of each have similar durations of $\sim 110$--120 days and subsequently drop by a comparable magnitude. Plateaus with edges this steep are not a typical characteristic of SNe~IIn. However, the plateau durations are very similar to normal SNe~II-P (Hamuy 2003). The subsequent decay tails of these SNe are relatively faint for SNe~IIn, but lie near the lower end of the range exhibited by SNe~II-P (Nadyozhin 2003; Smartt et al. 2009). The optical decline rates of SN~2011ht and SN~1994W are also relatively steep, at 0.03--0.05 mag\,day$^{-1}$. This behavior is rather uncharacteristic of normal SNe~II-P, perhaps more similar to H-deficient SNe~Ib/c, which typically exhibit faster decline rates of 0.01--0.02 mag\,day$^{-1}$ (Elmhamdi et al. 2011); still, SNe~2011ht and 1994W decline substantially more rapidly than even those objects, in the optical.

\begin{figure}
\includegraphics[width=3.3in]{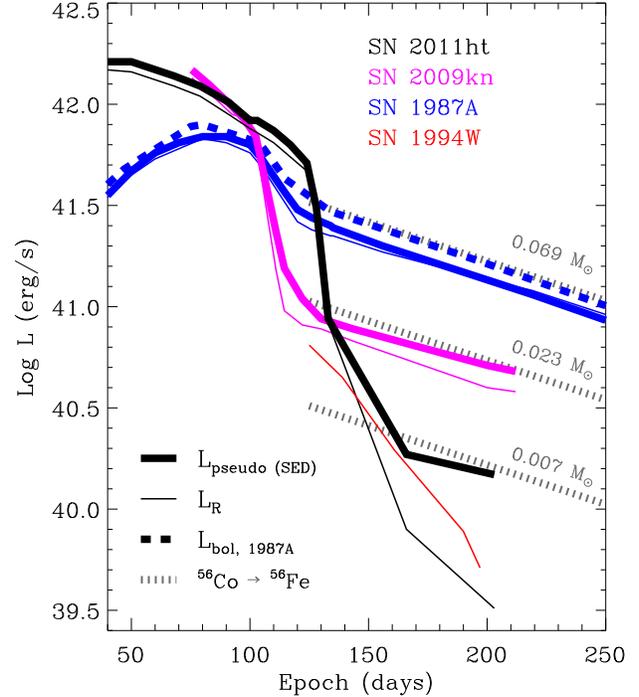}
\caption{Pseudo-bolometric (thick lines) and $R$-band (thin lines) luminosity light curves of SN~2011ht (black), SN~2009kn (red), and SN~1987A for comparison (blue). The pseudo-bolometric curves were derived by integrating the SEDs of both SNe between the $B$ and $K$ bands. For SN~1987A, this method underestimates the true bolometric luminosity by nearly 20\%, so we assume that this is a reasonable estimate for the error in our pseudo-bolometric curves. In the case of SN~2011ht, the slope of the late-time pseudo-bolometric tail is consistent with the rate of $^{56}$Co decay (dashed lines), even though the individual $R$-band tail is not. The result suggests that the unusually steep late-time decline in the optical could be the result of flux shifting to longer wavelengths at late times. The $^{56}$Co decay curves are shown for several values of $^{56}$Ni mass.}
\label{fig:mbol}
\end{figure}

We also derived a pseudo-bolometric luminosity for SN~2011ht. For the plateau phase we calculated bolometric corrections using our measured optical colors in conjunction with the method described by Bersten \& Hamuy (2009), which is valid only for the plateau. Using the extinction-corrected $B-V$ and $V-I$ colors, two independent values of pseudo-bolometric luminosity were computed for every epoch during the plateau and the results were averaged. The average difference between the two color-dependent luminosity values over all epochs was 0.36 dex. The rather large discrepancy probably results from the fact that this method was derived for SNe II-P, specifically, which generally have redder intrinsic colors than SNe IIn. Thus, we also calculated the pseudo-bolometric luminosity by integrating the source spectral energy distribution (SED) between the $B$ and $K$ bands, using trapezoidal integration. The resulting peak luminosity near day 55 is $L_{\rm peak}\approx2\times10^{42}$\,erg\,s$^{-1}$, which is slightly below the peak bolometric luminosity reported by Roming et al. (2012), probably because we have not accounted for the $U$-band optical and UV flux, which is significant. We then computed the total radiated energy of the plateau again using trapezoidal integration on the light curve, which yielded a total energy of $E_{\rm rad,p}\approx2\times10^{49}$ erg. Again, since this method did not account for the luminosity blueward of $B$ band and redward of $K$ band, it could be considered a lower limit. Nonetheless, the value is high, much higher than that of most SN impostors (Smith et al. 2011a), although normal for core-collapse SNe.

Although the post-plateau optical decline of SN~2011ht is unusually steep, the available $K$-band measurements during this phase (from Humphreys et al. 2012), also shown in Figure~\ref{fig:sn_lc}, appear to be consistent with the $^{56}$Co decay rate. The much smaller drop in brightness for the IR after the plateau suggests that a substantial amount of optical flux has shifted to longer wavelengths. So, to obtain a more accurate estimate of the post-plateau pseudo-bolometric luminosity of SN~2011ht, we, again, integrated their extinction-corrected SEDs between the $B$ and $K$ bands. To match the available IR post-plateau data for SN~2011ht from Humphreys et al. (2012) on day 166, we estimated the optical photometry of SN~2011ht by linearly interpolating between our measured points on days 141 and 203. For epoch 203, for which we have secure $VRI$ detections, we interpolated the $K$-band photometry along the line connecting the measurements on days 166 and 217 from Humphreys et al. (2012). The pseudo-bolometric curves of SN~2009kn were computed in the same way, adopting measurements from Kankare et al. (2012). Finally, we also performed the same SED integration for SN~1987A, using photometric data from the literature (Menzies et al. 1987; Catchpole et al. 1987, 1988), sampling only a small fraction of the available data to match our cadence of SN~2011ht; and we compared this result with the true bolometric luminosity curve from Bouchet et al. (1991).

The resulting pseudo-bolometric light curves of SN~2011ht, SN~2009kn, and SN~1987A are presented in Figure~\ref{fig:mbol}, along with the absolute $R$-band curves for comparison. Without a priori knowledge of the full SED, we are unable to quantify how much flux remains unaccounted for at wavelengths shorter than $B$ band or longer than $K$ band. Hence, the uncertainties in our luminosity calculations are difficult to ascertain. Some guidance may be provided by application of our SED-integration method to SN~1987A, whose true bolometric luminosity is well constrained. In this case, the true bolometric luminosities are only marginally higher than our pseudo-bolometric values during the plateau phase ($\approx$5\%), while they are $\approx$20\% higher during the post-plateau decay tail, likely as a result of the unaccounted flux outside of the $B$-band to $K$-band range we considered . Still, the shape of the light curves and the decline rate remains consistent between the true bolometric and pseudo-bolometric values of SN~1987A, which is reassuring. 

Although the \textit{optical} decay tail of SN~2011ht suggests a decline rate faster than the rate of $^{56}$Co decay, the pseudo-bolometric luminosity, which includes the near-IR flux, declines at a rate consistent with $^{56}$Co decay. The decay of tail SN~2011ht is relatively faint, however, compared with known SNe~IIn, perhaps because many of them continue strong CSM interaction into late phases. In any case, the late time luminosity of SN~2011ht lies near the lower end of the range exhibited by SNe~II-P (Nadyozhin 2003; Smartt et al. 2009), which suggests a relatively low mass of $^{56}$Ni. Overall, the behavior of SN~2011ht, like SN~2009kn, appears consistent with that of a bona fide core-collapse SN. 

\subsection{Spectral Evolution}
\begin{figure*}
\includegraphics[width=5.5in]{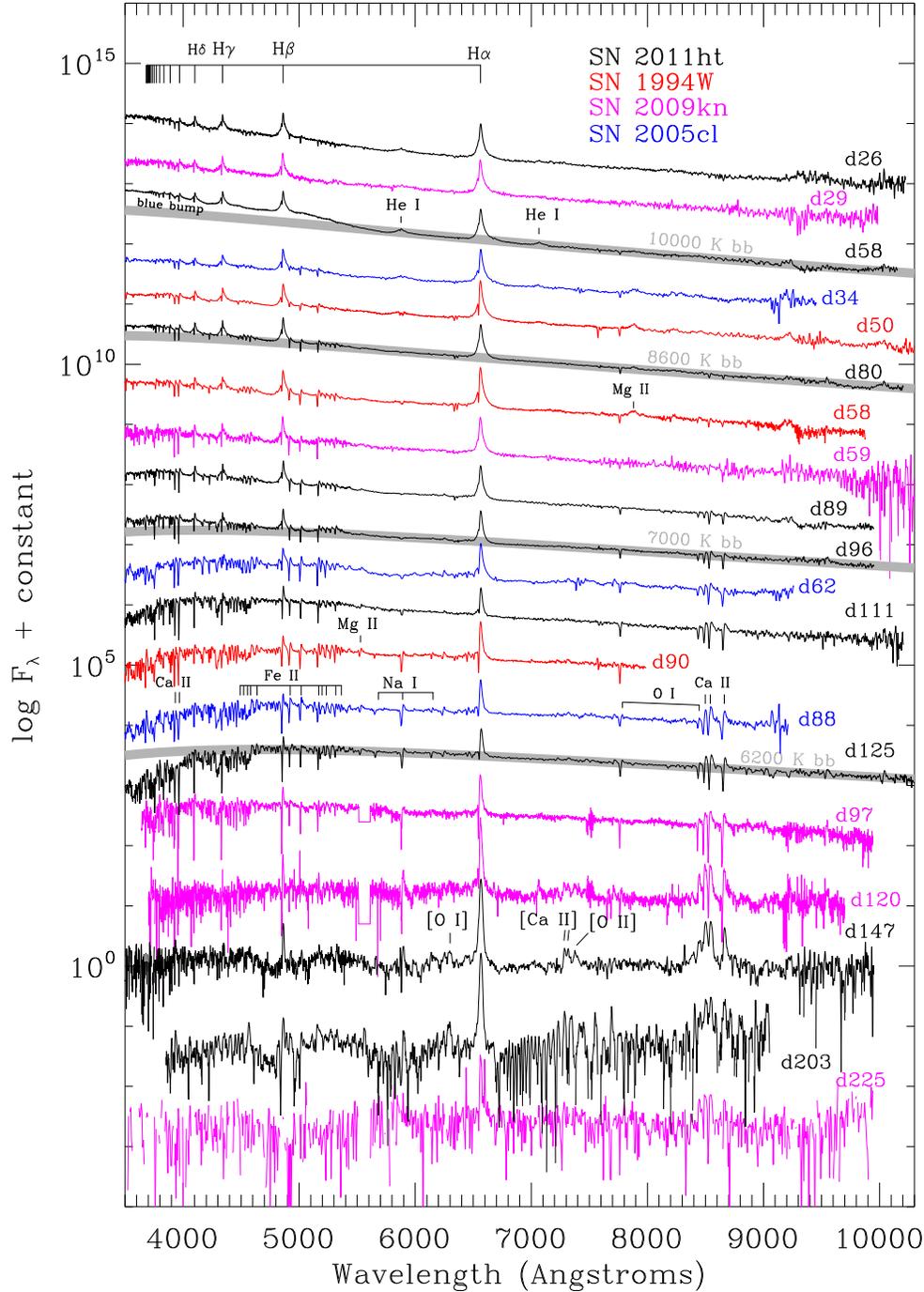}
\caption{Dereddened spectra of SN~2011ht and its kin: SN~1994W (\textit{red}), SN ~2009kn (\textit{magenta}), and SN~2005cl (\textit{blue}). The SN~2005cl spectra were obtained from the Weizmann interactive SN data repository (Yaron \& Gal-Yam 2012). Our own SN~2009kn spectra for days 29 and 59 are supplemented with spectra from days 97, 120, and 225 (Kankare et al. 2012), kindly provided by E. Kankare.}
\label{fig:lick_spec}
\end{figure*}

\begin{figure}
\includegraphics[width=3.3in]{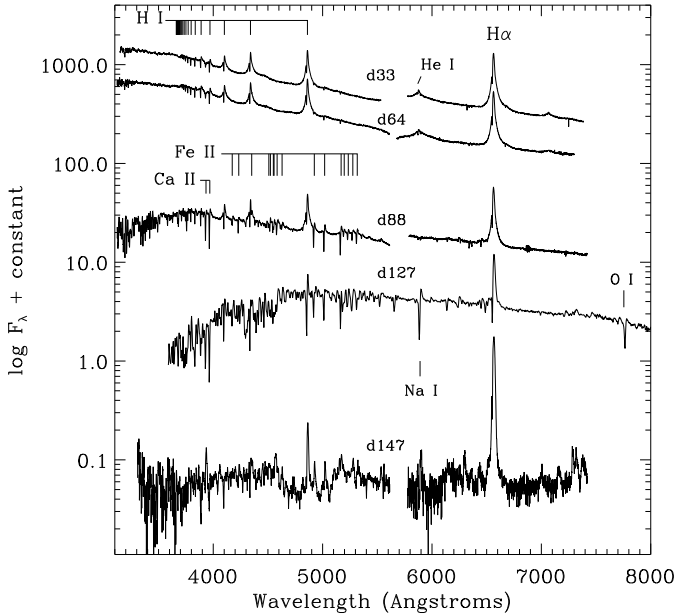}
\caption{Optical spectra of SN~2011ht from Keck/LRIS, and a KPNO/RCSpec spectrum from day 127.}
\label{fig:hires_spec}
\end{figure}

\begin{figure}
\includegraphics[width=3.3in]{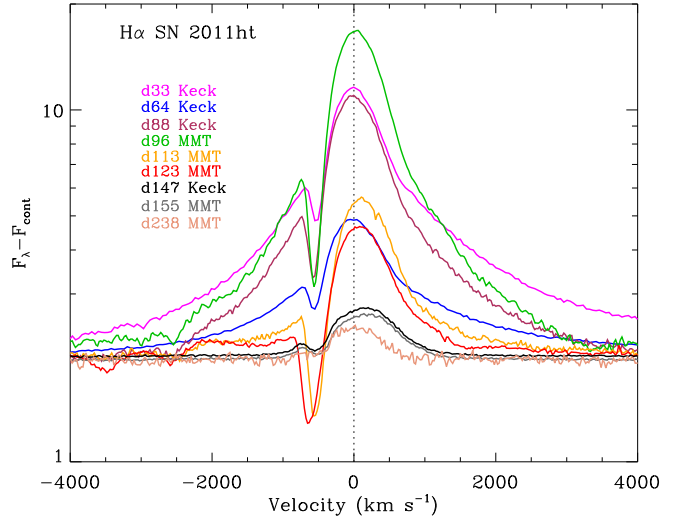}
\caption{Evolution of the continuum-subtracted H$\alpha$ profile of SN~2011ht between days 33 and 238 after discovery. At early times, the emission line exhibits an intermediate-width ($\sim 1500$\,km\,s$^{-1}$) core P-Cygni profile with superimposed broad Lorentzian wings. At later times, the wings diminish, yet P-Cygni absorption remains present well into the nebular phase.}
\label{fig:spec_ha}
\end{figure}

Our Lick/Kast spectra of SN~2011ht and SN~2009kn are presented in Figure~\ref{fig:lick_spec}, in addition to three later epochs of SN~2009kn from Kankare et al. (2012), and archival spectra of SN~1994W from Chugai et al. (2004). Spectra of SN~2005cl are also included for comparison, and were obtained from the Weizmann interactive SN data repository (Yaron \& Gal-Yam 2012). Figure~\ref{fig:hires_spec} shows our higher-resolution spectra from Keck/LRIS and KPNO/RCSpec. Figure~\ref{fig:spec_ha} displays the same Keck/LRIS spectra, in addition to our MMT spectra, all centered on the H$\alpha$ emission line. During our earliest epoch on day 26, SN~2011ht exhibits a characteristic SN~IIn emission-line spectrum, dominated by the Balmer series of H~{\sc i}. The emission lines exhibit intermediate-width cores of full width at half-maximum intensity (FWHM) $\sim 1500$\,km\,s$^{-1}$ on top of broad Lorentzian wings, which presumably form as a result of Thomson scattering of photons off of free electrons in the dense CSM and within the post-shock gas (Chugai et al. 2001; Dessart et al. 2009). The H$\alpha$ emission profile is slightly asymmetric, having a more extended red wing. The underlying broad components have full-width near zero intensity (FWZI) values of 8000--9000\,km\,s$^{-1}$ on day 33, and weaken to $\sim 7000$\,km\,s$^{-1}$ by day 64.  The lines exhibit narrow P-Cygni absorption components with blueshifted velocities of $\sim 500$--600\,km\,s$^{-1}$, which remain at roughly constant velocity throughout the entire plateau phase. Weak emission from He~{\sc i} $\lambda$5876 is present at the earliest epochs and strongest on day 58. Weak absorption features of Fe~{\sc ii} and Ca~{\sc ii} H\&K are present by day 80, in addition to weak Mg~{\sc ii} $\lambda$7888 in pure emission. By day 58 (when the light curve is near maximum) the spectrum has increased in ionization temperature, with the strengthening of He~{\sc i} $\lambda$5876 and the appearance of He~{\sc i} $\lambda$7065, in addition to the development of a strong blue excess that rises shortward of 5500\,\AA. The excess becomes apparent near a cluster of Fe~{\sc ii} lines, which become individually distinguishable as absorption-dominated P-Cygni features by day 80 and are accompanied by additional absorption features of this species in between H$\beta$ and H$\gamma$. Redward of this bump the slope of the continuum appears to be consistent with a 10,000\,K blackbody. 

The Balmer lines and Fe~{\sc ii} P-Cygni features appear in greater detail in our higher-resolution spectra (Fig.~\ref{fig:hires_spec}), revealing that the Fe~{\sc ii} lines also exhibit signs of Lorentzian broadening. Thus, the blue excess from day 58 appears to be the result of blended Fe~{\sc ii} emission features, like SN~2005ip (Smith et al. 2009) and SN~2006jc (Chugai 2009; Foley et al. 2007). 

On day 89 (for all SNe, the day is given with respect to the discovery date) the spectrum has begun to exhibit the characteristics of a cool, dense gas, very similar to SN~1994W on day 58 and SN~2009kn on day 59. Continuously strengthening Fe {\sc ii} features are accompanied by P-Cygni lines of the Ca~{\sc ii} triplet at 7500--8000\,\AA, all of which continue to increase in strength through days 125--127, at which point Na~{\sc i} P-Cygni and absorption features also become evident. By this time, the continuum temperature has decreased to $\sim 6200$\,K, and the broad Lorentzian wings of H$\alpha$ have become diminished beyond detectability, as evidenced by the higher-resolution spectra of H$\alpha$ shown in Figure~\ref{fig:spec_ha}.  In addition, line blanketing suppresses the flux at the blue end of the spectrum, which must be partially responsible for the greater drop in the $B$-band light curve in Figure~\ref{fig:sn_lc} (in addition to a decrease in blackbody temperature), before the plateau edge has been reached.  

By the time of the spectrum on day 147, the light curve exhibits a steady decline, the continuum has weakened substantially, and the P-Cygni profiles are less prominent. The higher-resolution H$\alpha$ spectrum in Figure~\ref{fig:spec_ha}, however, shows that P-Cygni absorption is still present. The Balmer series has weakened significantly, with only H$\alpha$, H$\beta$, and H$\gamma$ clearly visible. Forbidden transitions of [O~{\sc i}], [O~{\sc ii}], and [Ca~{\sc ii}] have also appeared, the latter of which implies very low density for the gas. These features imply that the SN has entered the nebular phase. All emission lines remain narrow ($< 1200$\,km\,s$^{-1}$) throughout our spectroscopic coverage of this phase.

Using the distance and extinction values that we adopted to generate the absolute-magnitude curve in Figure~\ref{fig:lc_abs}, we calculated the luminosities of the H$\alpha$ and H$\beta$ emission lines, and compared them with those of SN~1994W (data from Chugai et al. 2004) and SN~2009kn (Kankare et al. 2012). The same measurements were made for SN~2005cl, adopting the distance and extinction values from (Kiewe et al. 2012). The results are presented in Figure~\ref{fig:ha_lum}.  SN~1994W and SN~2005cl have a substantially higher H$\alpha$ luminosity than SN~2011ht and SN~2009kn during the plateau phase, by almost an order of magnitude, which could be the result of interaction with higher-density CSM, a different CSM filling factor (i.e., clumping), or geometric differences. After their plateau phases end around $\sim120$ days, the light curves of each SN converge to the same late-time H$\alpha$ luminosity. The H$\alpha$ decline rates of SN~2011ht and SN~1994W are identical throughout their nebular phases, and are steeper than SN~2009kn. Nebular phase coverage is not available for SN~2005cl, however. The fact that the line emission does not exhibit the same sharp plateau edge as the broad-band light curves of these SNe suggests that the continuum and H$\alpha$ emission source might not be coupled. 

After the plateau, H$\beta$ becomes noticeably fainter than H$\alpha$ for each SN, corresponding to a change in Balmer decrement. During the remainder of the nebular phase, H$\beta$ exhibits approximately steady decline, somewhat steeper than for H$\alpha$. 

\section{Discussion} 
SNe~2011ht, 1994W, and 2009kn share an unusual set of properties, particularly the combination of a luminous well-defined  plateau light curve, a faint decay tail, and nearly identical spectral evolution that is distinct from the larger class of SNe~IIn. SN~2005cl exhibits similar characteristics, although the lack of post-plateau information limits the comparison of this SN to the former group. The unique character of these SNe suggests that a very specific physical scenario has played out in each of these events, involving circumstellar interaction and low $^{56}$Ni yield. It therefore appears plausible that the same type of explosion has taken place in each case, whatever the underlying physics might be. Various possibilities, consistent with the available data, are discussed below.

\subsection{Circumstellar Interaction}
The photometric and spectroscopic evolution of SN~2011ht during the plateau phase, specifically the high luminosity and the relatively narrow P-Cygni emission-line profiles superimposed on the broad Lorentzian wings, indicates strong interaction with CSM. SN~1994W exhibited a nearly identical spectral morphology and evolution, and modeling of this SN by Chugai et al. (2004) demonstrated that homologous expansion of dense CSM is required to fit the observational data. Additional modeling of this SN by Dessart et al. (2009) resulted in different conclusions about the origin of the spectral features, but also found that CSM interaction produced a good match to the spectrum.  

The Chugai et al. (2004) model concluded that the CSM of SN~1994W consists of a 0.4\,M$_{\odot}$ envelope having an accelerated velocity gradient of 170--400\,km\,s$^{-1}$ within a radius of $3.3\times10^{15}$\,cm, which implies a very high mass-loss rate of 0.3\,M$_{\odot}$\,yr$^{-1}$ that probably occurred $\sim 1.5$\,yr before core collapse. Such a large mass-loss rate is beyond the physical parameters of a sustainable line-driven stellar wind, and implies an eruptive/explosive origin for the CSM (Smith \& Owocki 2006). The observational characteristics of SNe~2011ht and 2009kn, and perhaps SN~2005cl, are so similar to those of SN~1994W that one could reasonably arrive at a similar physical interpretation for these SNe. 

\begin{figure}
\includegraphics[width=3.3in]{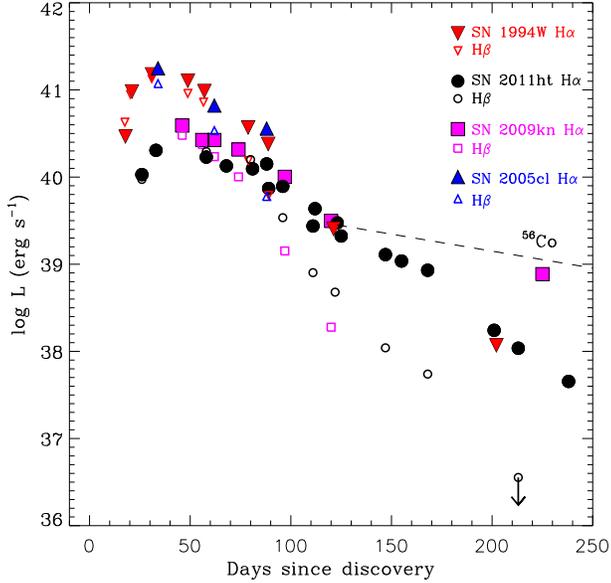}
\caption{Light curves of H$\alpha$ and H$\beta$ luminosity for SNe~2011ht, 1994W, 2009kn, and 2005cl. All objects exhibit very similar luminosity evolution and, in cases where late-time data are available, a steep late-time decline that is faster than the rate of $^{56}$Co decay. SN~1994W and 2005cl are relatively luminous during their plateau phases, although each object having late-time coverage converges to the same post-plateau H$\alpha$ luminosity around day 120. The late-time declines of SNe~2011ht and 1994W are identical, at least out to day 203.}
\label{fig:ha_lum}
\end{figure}

\begin{figure}
\includegraphics[width=3.3in]{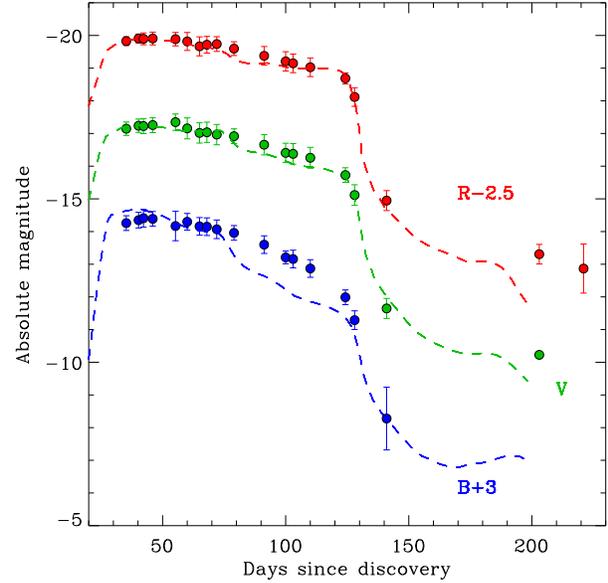}
\caption{$BVR$ light curves of SN~2011ht plotted against a scaled CSM-interaction light-curve model for the SN~IIn 1994W from Chugai et al. (2004).}
\label{fig:lc_wmod}
\end{figure}

Figure~\ref{fig:lc_wmod} shows the best-fit model light curves for SN~1994W from Chugai et al. (2004) --- specifically, their model ``sn94w43''. 
To match the SN~2011ht data, we shifted the SN~1994W models down by a factor of 6--7 in luminosity ($\sim 2$ mag) and forward in time by 16 days (perhaps accounting for uncertainties in the explosion dates).  For the $R$ and $V$ bands, the model plateau shape and the abrupt transition into the tail phase matches the SN~2011ht photometric data well, although the match to the $B$-band plateau data is not as satisfactory for the latter half of the plateau. 

The SN~2011ht data match the overall shape of the SN~1994W model light curve, but the factor of 6--7 discrepancy in peak H$\alpha$ luminosity must be accounted for.  Assuming strong interaction between a SN shock and surrounding CSM during the plateau phase, the luminosity of H$\alpha$ can be expressed as 
  \begin{equation}
L(\textrm{H}\alpha)\propto \frac{L_s}{T_s}\propto \frac{V_s^3}{T_s}\left ( \frac{\dot{M}}{v_w} \right),
 \end{equation} 
where $L_s$, $T_s$, and $V_s$ are the shock ionizing luminosity, temperature, and velocity (respectively), and $w=\dot{M}/v_w$ is the wind-density parameter. For SN~1994W, Chugai et al. (2004) estimated the shock velocity to be 4000\,km\,s$^{-1}$ and the shock temperature to be within a reasonable range of 1--$2\times10^4$\,K. Their model, which requires a 0.3\,M$_{\odot}$\,yr$^{-1}$ mass-loss rate and 400\,km\,s$^{-1}$ wind velocity, implies a very high wind-density parameter. Because of the high sensitivity of H$\alpha$ emission to shock velocity, a factor of less than 2 difference in $V_s$ alone could account for the factor of 6--7 lower luminosity of H$\alpha$, assuming the CSM wind-density parameters were similar for SN~2011ht and SN~1994W.  This could be achievable by either an intrinsically faster shock, or a bigger velocity difference between the shock and the expanding CSM. Alternatively, if the shock velocities (and temperatures) are the same, which is supported by the very similar plateau durations, then the luminosity differences could instead indicate a lower-density CSM, or a lower geometric covering factor (solid angle) associated with the CSM of SN~2011ht. Assuming that differences in the CSM velocity between SN~2011ht and SN~1994W could be traced approximately by the relative blueshifts of their H$\alpha$ P-Cygni absorption components, then the factor of 1.5 smaller blueshift for SN~2011ht would imply a factor of $\sim 1.5$ \textit{increase} in H$\alpha$ luminosity. The fact that H$\alpha$ is nearly an order of magnitude fainter for SN~2011ht during the plateau thus requires an order of magnitude lower mass-loss rate for the progenitor, or 0.03\,M$_{\odot}$\,yr$^{-1}$.  This value is still extremely high, however, and thus would also imply an eruptive/explosive origin for the CSM associated with SN~2011ht. 

The luminosity and decline rate of H$\alpha$ during the nebular phase is practically identical for SN~2011ht and SN~1994W. In the latter case, Chugai et al. (2004) used the late-time H$\alpha$ luminosity to constrain the outer pre-eruption wind parameters of the progenitor. The low luminosity of SN~1994W on day 203 ($L\approx10^{38}$\,erg\,s$^{-1}$)  implies that the outer wind should have a density $\sim 10$ and $\sim 2$ times lower than that of the well-studied SN 1979C and SN 1980K (respectively), which presumably had red-supergiant progenitors (Fesen \& Becker 1990; Weiler et al. 1991). The dramatic change in mass-loss rate supports the hypothesis of eruptive/explosive mass loss several years preceding the SN. Again, owing to the striking similarities shared with SN~1994W, both qualitatively and quantitatively, one arrives at a similar conclusion for the outer-wind parameters of SN~2011ht. 

The very steep plateau edges in the light curves of SNe~2011ht, 1994W, 2009kn, and 2005cl are rather unusual for SNe~IIn, which normally exhibit a slower and more steady decline. In the case of SN~1994W, Chugai et al. (2004) interpreted this characteristic as the result of the SN shock wave reaching the edge of a dense CSM envelope, while Dessart et al. (2009) favored simple photospheric contraction as a result of hydrogen cooling to recombination temperature. The latter scenario was also favored by Kankare et al. (2012) in the case of SN~2009kn. However, both of these scenarios could play a role in producing the sharp plateau edge. As the SN shock propagates through the dense CSM, the cool dense shell (CDS) that develops at the SN-CSM interface will create an optically thick barrier that masks the inner high-velocity ejecta. Once the presumed ``edge" of the dense CSM envelope is reached, the CDS should rapidly expand and become optically thin. By this time, if the inner ejecta have already cooled below the recombination temperature of hydrogen ($\sim 6000$\,K), the explosion should quickly become transparent to the observer, resulting in a rapid drop in continuum flux, and a fast transition into the nebular phase. Since the radiative energy of the inner ejecta is dominated by the thermalization of radioactively generated photons, a sufficiently low mass of radioactive $^{56}$Ni could potentially result in such a scenario by allowing the interior to rapidly cool and become transparent before the CDS reaches the edge of the CSM and thins out. Low $^{56}$Ni mass is also consistent with the small velocity widths of the emission lines observed during the nebular phases of SNe~2011ht, 1994W, and 2009kn (Maguire et al. 2012).  We demonstrate below that the late-time luminosity of SN~2011ht also implies a low mass of radioactive $^{56}$Ni synthesized in the explosion.

\subsection{Low $^{56}$Ni Mass}
The decay tails of SNe~2011ht, 1994W, and 2009kn are faint and imply a relatively low yield of $^{56}$Ni compared with other known SNe~II (Smartt et al. 2009).  Since radioactivity provides thermal energy to the ejecta, a larger synthesized mass of $^{56}$Ni will not only result in a brighter decay tail in the late-time light curve, but also extend the hydrogen recombination time scale, resulting in a shallower plateau edge and smoother transition into the nebular phase, as exhibited by the light curve of SN 1987A included in Figure~\ref{fig:lc_abs}. SNe that yield a small $^{56}$Ni mass exhibit the faintest decay tails and the sharpest plateau edges (e.g., see Elmhamdi et al. 2003). Thus, SNe~2011ht, 1994W, and 2009kn appear fully consistent with relatively low $^{56}$Ni-yield core-collapse explosions in all respects.

According to Sutherland \& Wheeler (1984), the relationship between the bolometric luminosity of the decay tail and the total mass of $^{56}$Ni can be expressed as 
  \begin{equation}
L=1.42\times10^{43}\,\textrm{erg\,s}^{-1}~\textrm{e}^{-t/111~\textrm{\tiny{days}}}~M_{\textrm{\tiny{Ni}}}/{\rm M}_{\odot},
 \end{equation} 
assuming 100\% $\gamma$-ray deposition efficiency into the SN envelope. For the moment, let us assume that the light-curve tail of SN~2011ht is dominated by $^{56}$Co decay, ignoring potential contribution from residual CSM interaction at late times. The pseudo-bolometric values we derived via SED integration imply a total luminosity of $L_{\rm bol}=1.9\times10^{40}$\,erg\,s$^{-1}$ for SN~2011ht on day 166, recalling that the SED was constructed using the the IR measurements from Humphreys et al. (2012) on day 166  with our interpolated optical photometry on the same day. This would imply a $^{56}$Ni mass of $\sim0.006$\,M$_{\odot}$ using Equation~2. 

The uncertainty in our $^{56}$Ni mass estimate is difficult to ascertain, since we interpolated the optical light curve between our measurements on days 141 and 203. Furthermore, as indicated by our analysis on SN~1987A, which we performed as a check on our method for measuring the pseudo-bolometric luminosity via SED integration, the results underestimated the true bolometric luminosity by $\sim$20\% during the nebular phase, which would imply that the $^{56}$Ni mass of SN~2011ht is probably closer to 0.007--0.008~M$_{\odot}$. However, this reasoning does not account for potential differences between the SEDs of SN~2011ht and SN~1987A.  Finally, residual emission from CSM interaction could contribute to the late-time luminosity. Thus, we conclude conservatively that the $^{56}$Ni mass of SN~2011ht  lies in the range of 0.006--0.01 $M_{\odot}$. Despite the large uncertainty, we can state with confidence that the range of likely values for the $^{56}$Ni mass occupies the low end of the range exhibited by SNe~II-P (Turatto et al. 1990; Sollerman 2002; Nadyozhin 2003; Smartt et al. 2009), which has important implications that will be discussed below. 

One can also estimate the $^{56}$Ni mass by direct comparison of decay-tail luminosity to that of the nearby SN 1987A, which provides a reliable benchmark, owing to its well-constrained distance and luminosity. The mass of $^{56}$Ni produced by SN 1987A is $\sim0.07$\,M$_{\odot}$ (Suntzeff \& Bouchet 1990). The luminosity of SN 1987A on day 203 was $1.6\times10^{41}$\,erg\,s$^{-1}$ (Suntzeff et al. 1991). By direct comparison, the luminosity ratio of  0.09 also implies an initial $^{56}$Ni mass of $0.006$\,M$_{\odot}$ for SN~2011ht (using the pseudo-bolometric luminosity), consistent with the value from the method of Sutherland \& Wheeler (1984). The SN 1987A comparison method was used by Kankare et  al. (2012) to estimate a value of 0.023\,M$_{\odot}$ for SN~2009kn, which was regarded as a lower limit, since the shallower decline rate for this SN indicates the likelihood that persistent CSM interaction continued well into the nebular phase and thus contributed to the late-time bolometric luminosity. Our analysis on SN~2009kn data yields a $^{56}$Ni mass consistent with the results of Kankare et al. (2012). Note that in this analysis we assumed the explosion date is the discovery date. We expect that the uncertainty in explosion date for SN~2011ht is likely to be less than 1--2 weeks, since the early UV rise was observed by Roming et al. (2012). A two-week difference would raise the derived $^{56}$Ni mass by only 14\%, which does not change our conclusion that the $^{56}$Ni yield of SN~2011ht is relatively low compared with other core-collapse SNe.

Had we not considered the total optical-IR SED luminosity of SN~2011ht, and instead only used the optical photometry, we would have significantly underestimated the luminosity and, hence, the $^{56}$Ni mass. For example, using its absolute magnitude of $M_V=-10.67$ mag on day 203 and adopting a bolometric correction of $-0.45$ mag for the nebular phase (Bersten \& Hamuy 2009), the  total luminosity calculated for day 203 would be $5.7\times10^{39}$\,erg\,s$^{-1}$, implying a factor of 4--5 lower $^{56}$Ni mass of $2.4\times10^{-3}$\,M$_{\odot}$. The underestimated value matches the \textit{lower} end of the range estimated for SN~1994W by Sollerman et al. (1998), who did not have IR measurements at their disposal. Thus, the true $^{56}$Ni mass of SN~1994W, if its infrared decay tail properties are similar to SN~2011ht, could be closer to the \textit{upper} end of their derived range (0.015\,M$_{\odot}$), for which they considered dust as a potential cause for the unusually steep late-time decline of the optical light curve.

\subsection{The Possible Influence of Dust}

For SN~1994W, dust formation could have been more tightly constrained if multi-band optical and IR coverage had been obtained at late times. For SN~2011ht, we were fortunate to have obtained multi-band optical measurements during the nebular phase, which can be combined with the late-time infrared measurements of SN~2011ht reported by Humphreys et al. (2012). The IR data, included in Figure~\ref{fig:sn_lc}, unambiguously demonstrate that substantial flux has shifted into the IR after the plateau phase, and this has been attributed to the formation of dust. However, if the steep optical decline were caused by extinction from such dust, then we might have expected substantial reddening to be observed in the $V-R$ and $V-I$ colors between days 141 and 203. As shown in Figure~\ref{fig:sn_lc}, no such reddening is observed after the plateau phase ends. In fact, the colors appear to flatten out and become slightly bluer, probably because the continuum during the end of the plateau, which exhibited strong line blanketing at the blue end of the spectrum, has dropped out almost completely by day 147. 

However, the photometric data are perhaps a bit too sparse at late times to justify making any definitive conclusions about the influence of dust and increasing extinction. However, the late-time luminosity ratio of H$\alpha$ and H$\beta$ emission, shown in Figure~\ref{fig:ha_lum}, does appear to be consistent with increasing reddening, as the last H$\beta$ measurement after day 200 deviates from H$\alpha$ more extremely than for earlier epochs. Alternatively, in the optically thin regime, collisional excitation can result in significant deviations from Case B recombination, which will also result in a decreasing H$\beta$ luminosity; this could be important for the nebular phase. We note that if extinction by dust is important for H$\alpha$, then the estimated CSM wind-density parameters that are based on the luminosity of this line could be significantly underestimated for SN~2011ht. 

In conclusion, although the cause of the steep optical decline of SN~2011ht during the nebular decay tail remains unclear, the hypothesis of dust formation is hard to rule out. The steep optical decline during the nebular phase could likely to be the result of the absorption and thermal reprocessing of photons. Such dust could have condensed following the steep drop in luminosity after the plateau, although the rapid time scale over which grains would have had to grow is difficult to reconcile (Humphreys et al. 2012). Whatever the case, the properties of SNe~2011ht, 1994W, and 2009kn underscore the importance of obtaining IR photometry to accompany optical measurements during both the plateau phase and later nebular phases, so that the full bolometric luminosity can be precisely estimated and the possible influence of dust can realized.

\subsection{The Nature of the Explosion and the Progenitor}
The results of our observations and analysis suggest a low $^{56}$Ni mass of 0.006--0.01\,M$_{\odot}$ for SN~2011ht, and this has important implications for the nature of the progenitor. $^{56}$Ni yields this low could result from the lowest-mass stars believed to be capable of undergoing core collapse. Stars with initial masses of  8--10\,M$_{\odot}$ are not thought to produce an Fe core. Instead, they are thought to end their lives as super-asymptotic-giant-branch (SAGB) stars which develop electron-degenerate O-Ne-Mg cores (Barkat 1974; Nomoto 1984). The favored core-collapse mechanism in this case is electron capture onto $^{24}$Mg and $^{20}$Ne, which triggers collapse before the onset of explosive O burning (Miyaji et al. 1980).  Owing to the low overall mass and the neutron richness of the ejecta, nucleosynthesis models for electron-capture SNe predict low $^{56}$Ni yields, perhaps as small as  0.002\,M$_{\odot}$ (Wanajo et al. 2009), but more conservatively  $<0.015$\,M$_{\odot}$ (Kitaura et al. 2006). The model yields for 8--10~$M_{\odot}$ stars appear to be consistent with the volume-limited sample of relatively nearby core-collapse SNe that have identified progenitors and constrained initial masses (Smartt et al. 2009).

The light curves of  SN~2011ht, 1994W, 2009kn, and 2005cl bear some resemblance to other interacting SNe which have been considered viable candidates for electron-capture SNe. The low-luminosity SN~2005cs, whose light curve is shown in Figure~\ref{fig:lc_abs}, exhibited a similarly steep plateau edge and a subluminous decline, implying a very low $^{56}$Ni mass of $3\times10^{-3}\,{\rm M}_{\odot}$ (Pastorello et al. 2009). The progenitor had an estimated mass of $9_{-2}^{+3}\,{\rm M}_{\odot}$, derived from pre-SN images of its nearby host galaxy M51 (Maund et al. 2005). Based on these properties, SN~2005cs was considered a viable candidate for an electron-capture SN. However, the IR colors of the progenitor derived from archival \textit{HST} images, appeared inconsistent with those of an SAGB star (Eldridge et al. 2007), which is the expected progenitor of such an explosion. 

Other candidates for electron-capture SNe from SAGB progenitors are SN~2008S (Prieto et al. 2008; Botticella et al. 2009) and SN~2007od (Inserra et al. 2011). SN~2008S exhibited a spectral morphology and evolution similar to SNe~2011ht, 1994W, and 2009kn, indicative of CSM interaction. Moreover, SED modeling of its strong mid-IR excess indicated that the CSM likely consists of multiple dusty optically thick shells. Although the light curve plateau of SN~2008S did not exhibit as steep of an edge as SN~2011ht, the derived $^{56}$Ni mass was nonetheless a very low $\sim$10$^{-3}$ $M_{\odot}$. Evidence in support of an SAGB progenitor for SN~2008S is also strong in this case, owing to the detection of the progenitor star in archival mid-infrared images from the \textit{Spitzer Space Telescope}, which exhibited an SED consistent with this interpretation (Thompson et al. 2009; Prieto et al. 2009).  Finally, the light curve of SN~2007od exhibited a well-defined plateau with a duration of 100--120 days, similar to that of SN~2011ht. However, this object's spectral evolution during the plateau exhibited very broad lines typical of normal SNe II-P, in contrast to the narrow-lined spectra exhibited by SN~2011ht and SN~2008S. Although evidence for CSM interaction did appear in the late spectra of SN~2007od, in the form of a boxy multi-peaked H$\alpha$ line (Inserra et al. 2011), the persistence of broad lines throughout its entire evolution imply that an optically-thick shell capable of masking the inner high-velocity SN ejecta never developed, perhaps because the CSM mass and density was relatively low, or because of an aspherical geometry for the CSM. 

It seems plausible that SN~2011ht could be an explosion from an 8--10 $M_{\odot}$ progenitor, as already suggest for SN~2009kn by Kankare et al. (2012), although with more luminous CSM interaction. The nebular phase spectra might also be consistent with this interpretation. Specifically, the feeble emission from [O~{\sc i}] $\lambda$6303, which is weak in SN~2011ht and not detected in SN~2009kn or SN~1994W, favors a relatively low-mass progenitor, as indicated by Dessart et al. (2010).   

If SNe~2011ht, 1994W, and 2009kn did arise from 8--10 $M_{\odot}$ progenitors, then the minor differences between their photometric and spectroscopic evolution could potentially be the result of variations within the masses and radial density profiles of their CSM envelopes, in addition to differences in explosion energy. In the case of SN~1994W, which we believe has CSM parameters similar to SN~2011ht,  Chugai et al. (2001) estimated a CSM mass of 0.3 $M_{\odot}$ with a radius of $R\sim300$AU, and an approximate volume density of $n\sim1\times10^9$ cm$^{-3}$. In the case of SN~2008S, SED modeling indicates a significantly lower mass and density CSM of $M=10^{-3} M_{\odot}$ and $n=3\times10^{7}$ cm$^{-3}$ (Prieto et al. 2008), which results in less efficient conversion of SN kinetic energy into optical light, and, hence, a lower peak luminosity for the light curve. 

Indeed, the CSM configurations of electron-capture progenitor AGB stars are potentially very diverse, exhibiting a large variety at the time of core collapse. Depending on factors such as metallicity  (Pumo et al. 2009), possible activity from He or C burning flashes, or perhaps core-O or Ne flashes. For SN~2011ht, the ejection of CSM in the several years before core collapse could have resulted from explosive instabilities that occur during the late nuclear burning stages and/or SAGB winds (e.g., Poelarends et al. 2008, and references therein). However, the 500--800\,km\,s$^{-1}$ velocity of the P-Cygni absorption components of SNe~2011ht  are significantly faster than the observed wind velocities of SAGB stars ($\sim 10$\,km\,s$^{-1}$). Still, there is no particular reason to suspect that eruptive pre-SN mass loss from final nuclear flashes should share the low outflow velocities of earlier and more steady SAGB winds. 

As an alternative to the possibility of low-mass progenitors and electron-capture SNe, a low $^{56}$Ni yield explosion can result if a substantial fraction of the inner core ejecta does not achieve the escape velocity necessary for a successful explosion and thus falls back onto the compact remnant, perhaps accreting onto a neutron star or descending into a black hole (Zampieri et al. 2003). If this is the case, then the progenitors of SN~2011ht and its kin could have been substantially more massive than 8--10\,M$_{\odot}$, perhaps having initial masses of 30\,M$_{\odot}$ or more (Fryer 1999). Consider the case of the jet-powered SN 2010jp, which has been suggested to possibly mark the formation of a black hole (Smith et al. 2012b). This SN~IIn exhibited a steeply declining decay tail with a low overall luminosity, which constrained the $^{56}$Ni mass to $< 0.003\,{\rm M}_{\odot}$. Black hole formation is likely preceded by substantial fallback of SN ejecta, so low $^{56}$Ni masses are to be expected in such a scenario. 

There has also been the suggestion that SN~2011ht may represent a non-terminal event. Dessart et al. (2009) suggested that SN~1994W could have resulted from the interaction of colliding circumstellar shells that were ejected from consecutive mass eruptions, where the more recent ejection has a higher velocity and catches up with the slower outer shell, creating a luminous collision by conversion of kinetic energy into UV-optical-IR light. If SN~2011ht was the result of an LBV eruption, however, it would be a rather remarkable set of coincidences for the end of the plateau phase to occur at $\sim 120$ days, which is a very typical time scale for SNe~II-P, and for the late-time bolometric decline to be consistent with the $^{56}$Co decay rate. The uncanny spectroscopic and photometric similarity of SN~2011ht and SN~1994W to SN~2009kn and SN~2005cl, which were almost certainly  bona fide core-collapse events, would also be a rather remarkable coincidence.    

Arguments in favor of the nonterminal LBV scenario (e.g., Humphreys et al. 2012) are based on a plausibility argument, and thus provide insufficient evidence against core collapse. The lack of observational signatures from high-velocity material in the spectra of SNe~2011ht, 1994W, and 2009kn can be explained by the presence of an optically thick CDS that remains present during the entire plateau phase, which masked emission from inner high-velocity material. A similar process was invoked to explain the spectrum of the luminous SN~2006gy (Smith et al. 2010), which shares many spectral similarities with these SNe. By the time the CDS thinned out, the explosion could have cooled and become transparent, never revealing the observational signature of high-velocity ejecta. As we pointed out earlier, the exceptionally low $^{56}$Ni masses derived for SN~2011ht, SN~1994W, and, to a lesser extent, SN~2009kn, could allow for such rapid cooling and optical thinning of the ejecta, as the radioactive energy source was weak in these cases compared with normal SNe~II.  Finally, the interpretation of the SN~2011ht spectrum as the result of a radiatively driven ``opaque wind" is not physically plausible. Assuming that Thompson scattering dominates the optical depth ($\kappa_e=0.34$),  the SN~2011ht  peak luminosity of $L_{\rm peak}=(5\pm2)\times10^{42}$\,erg\,s$^{-1}$ would imply an Eddington factor of $\Gamma = (\kappa_e L)/4 \pi GMc \approx 1000$, which would undoubtedly result in explosive mass loss, not a wind, as suggested by Humphreys et al. (2012).

The only reliable means to observationally discriminate between the possibility of electron-capture SNe, a fallback scenario, or surviving LBVs would be the detection of the progenitor from deep pre-SN space-based images of the host galaxy, or an upper limit in the case of the electron-capture scenario. Unfortunately, pre-existing SDSS images of the SN~2011ht host, UGC 4560, do not place firm constraints on the nature of the progenitor, other than the fact that it could not be a progenitor as luminous as $\eta$ Car (Roming et al. 2012). However, LBVs on the fainter end of the distribution (Smith, Vink, \& de Koter 2004) may be possible. Still, scenarios that produce multiple massive shell ejections, such as the pulsational pair instability (Woosley et al. 2007), require extremely massive stars ($>90\,{\rm M}_{\odot}$), which would also be extremely luminous and can therefore be ruled out. This scenario would also seem unlikely to produce three SNe having such similar observational characteristics.  

\section{Conclusions}
We have presented extensive photometric and spectroscopic coverage of SN~2011ht. Our data complement those published by Roming et al. (2012) and Humphreys et al. (2012), but we arrive at somewhat different conclusions. This SN~IIn shares with SN~1994W and SN~2009kn an unusual combination of properties, including a well-defined and abruptly ending plateau phase dominated by CSM interaction, and low yields of $^{56}$Ni. The results of our analysis, particularly the late-time bolometric decline being consistent with the $^{56}$Co decay rate, supports the hypothesis that SN~2011ht, like SN~2009kn, was a core-collapse SN, and not the result of a super-Eddington wind. The more exotic scenario of multiple colliding LBV shells (Dessart et al. 2009) is a little more difficult to firmly rule out, although there is no reason why such a scenario should produce the plateau duration of $\sim 120$ days, which is so common among SNe~II-P (Hamuy 2003), or a late-time decline that is consistent with $^{56}$Co decay. The most likely scenario is that SN~2011ht was a bona fide core-collapse SN that produced a low $^{56}$Ni yield and exploded into dense CSM. Its discovery marks the third or fourth addition to an unusual group of SNe~IIn, including SN~1994W and SN~2009kn, and perhaps SN~2005cl. We argue that this justifies the designation of a new subclass of SNe exhibiting Type~IIn spectra, well-defined plateau light curves, and faint decay tails, which we propose be Type IIn-P.  The low synthesized mass of $^{56}$Ni for these events highlights the possibility that they are either the result of electron-capture SNe from 8--10\,M$_{\odot}$ progenitors, or explosions which experienced substantial fallback of their inner metal-rich ejecta. Additional members of the Type IIn-P class will undoubtedly emerge from future all-sky surveys, such as Pan-STARRS, the Large Synoptic Survey Telescope (LSST), and SkyMapper, and will enable the physical nature of these explosions to be elucidated more completely. Statistics for their rates and host-galaxy environments may also help discriminate between the two likely origins.

\section*{Acknowledgments}
This work is based in part on observations made at the the MMT Observatory, a joint facility of the Smithsonian Institution and the University of Arizona, and at the Lick Observatory, which is owned and operated by the University of California. Some of the data presented herein were obtained at the W. M. Keck Observatory, which is operated as a scientific partnership among the California Institute of Technology, the University of California, and NASA; the observatory was made possible by the generous financial support of the W. M. Keck Foundation. Infrared data herein were obtained using PAIRITEL, which is operated by the Smithsonian Astrophysical Observatory (SAO) and was made possible by a grant from the Harvard University Milton Fund, the camera loan from the University of Virginia, and the continued support of the SAO and UC Berkeley. We thank the staffs at these observatories for their efficient assistance. We are grateful to P. E. Nugent, D. Cohen, B. Y. Choi, M. Ellison, M. Mason, A. Wilkins, P. Blanchard, M. T. Kandrashoff, and O. Nayak for their help with observations at Keck and Lick. We also thank A. A. Miller, D. Starr, and C. Blake for facilitating PAIRITEL data contributions. E. Kankare kindly provided several epochs of spectra of SN~2009kn for our comparison. The supernova research of A.V.F.'s group at U.C. Berkeley is supported by Gary \& Cynthia Bengier, the Richard \& Rhoda Goldman Fund, the Christopher R. Redlich Fund, the TABASGO Foundation, and NSF grants AST-0908886 and AST-1211916.

\end{document}